\documentclass{sigchi}
\newcommand{\rev}[1]{{\color{black} {#1}}}

\toappear{Accepted for presentation at CSCW'16}

\usepackage{balance}  
\usepackage{graphics} 
\usepackage{txfonts}
\usepackage{times}    
\usepackage{url}      
\usepackage{color}
\usepackage{textcomp}
\usepackage{booktabs}
\usepackage{subfig}
\usepackage{dcolumn}
\usepackage{pifont}

\makeatletter
\def\url@leostyle{%
  \@ifundefined{selectfont}{\def\UrlFont{\sf}}{\def\UrlFont{\small\bf\ttfamily}}}
\makeatother
\def\pprw{8.5in}
\def\pprh{11in}

\usepackage[pdftex]{hyperref}
\definecolor{linkColor}{RGB}{6,125,233}
\hypersetup{%
  pdftitle={SIGCHI Conference Proceedings Format},
  pdfauthor={LaTeX},
  pdfkeywords={SIGCHI, proceedings, archival format},
  bookmarksnumbered,
  pdfstartview={FitH},
  colorlinks,
  citecolor=black,
  filecolor=black,
  linkcolor=black,
  urlcolor=linkColor,
  breaklinks=true,
}

\newcommand\tabhead[1]{\small\textbf{#1}}

\newcommand*{\sss}{\rlap{$^{***}$}}
\renewcommand*{\ss}{\rlap{$^{**}$}}
\newcommand*{\s}{\rlap{$^{*}$}}

\newcommand{\T}{True}
\newcommand{\F}{False}

\begin{document}


\title{Style in the Age of Instagram}
\subtitle{Predicting Success within the Fashion Industry using Social Media}

\author{}

\numberofauthors{3}
\author{%
  \alignauthor{Jaehyuk Park\\
    \affaddr{Indiana University}\\
    \affaddr{Bloomington, IN, USA}\\
    \email{jp70@indiana.edu}}\\
  \alignauthor{Giovanni Luca Ciampaglia\\
    \affaddr{Indiana University}\\
    \affaddr{Bloomington, IN, USA}\\
    \email{gciampag@indiana.edu}}\\
  \alignauthor{Emilio Ferrara\\
    \affaddr{Indiana University}\\
    \affaddr{Bloomington, IN, USA}\\
    \email{ferrarae@indiana.edu}}\\
}

\maketitle


\begin{abstract}

  Fashion is a multi-billion dollar industry with social and economic
  implications worldwide. To gain popularity, brands want to be represented by
  the top popular models. As new faces are selected using stringent (and often
  criticized) aesthetic criteria, \emph{a priori} predictions are made difficult
  by information cascades and other fundamental trend-setting mechanisms.
  However, the increasing usage of social media within and without the industry
  may be affecting this traditional system. We therefore seek to understand the
  ingredients of success of fashion models in the age of Instagram. Combining
  data from a comprehensive online fashion database and the popular mobile
  image-sharing platform, we apply a machine learning framework to predict the
  tenure of a cohort of new faces for the 2015 Spring\,/\,Summer season
  throughout the subsequent 2015-16 Fall\,/\,Winter season. Our framework
  successfully predicts most of the new popular models who appeared in 2015. In
  particular, we find that a strong social media presence may be more important
  than being under contract with a top agency, or than the aesthetic standards
  sought after by the industry.

\end{abstract}


\category{}{Human-centered computing}{Collaborative and social computing}[Social media]
\category{}{Information systems}{World Wide Web}[Social networks]

%
%
%

\section{Introduction}

The success of cultural artifacts is characterized by inherent unpredictability
and inequality~\cite{Salganik2006}, posing fundamental problems for a proper
understanding of markets based on the production of cultural goods. Fashion, and
fashion modeling in particular, are typical examples~\cite{Aspers2013,
Mears2011}. When trying to cast a model for the upcoming seasons, a casting
director is faced with a seemingly impossible task: predicting whom, out of the
hundreds of new faces she may see at the go-see calls, will become the top model
of the next season.

Modeling has in fact a very special meaning in fashion, a multi-billion dollar
industry with strong social and economical implications worldwide
\cite{Kawamura2004}. Models play the main role in advertisements and runways,
which are, historically, the main ways brands communicate with their customers.
They contribute to frame consumer experience and promote consumption, as their
attractiveness becomes associated to the brands they work
for~\cite{wissinger2009modeling}. This is especially true for luxury goods,
whose aesthetic value is more important than practical usage. Only those
who appeal the aesthetic sensibility of fashion designers stand
chances to become popular~\cite{djelic1999coevolution}.

Ethnographic studies show in fact that casting directors consider both objective
physical characteristics --- such as body size, height --- and
subjective considerations --- the reputation of the agency representing the
model --- to be important decision-making criteria for
casting~\cite{mears2005not, godart2009cultural, Mears2011}. However, the same
studies also uncover how information cascades are a critical part of what makes
a model successful. Similarly to the careers of scientists \cite{Merton1988},
fashion models also benefit from strong cumulative advantage effects, by which
small differences in prestige between competing individuals get amplified, for
example by means of word of mouth
\cite{Petersen2014,Rijt2014,godart2009cultural}.

The job market for fashion castings has strong seasonal components, revolving
around week-long trend-setting industry events (``Fashion Weeks''), during which
a dense calendar of shows is organized in various locations in a major city. The
four most prominent Fashion Weeks worldwide take place twice a year and, as of
2015, are hosted in the cities of New York, London, Paris, and Milan. These
events facilitate networking and information sharing, and are seen as a crucial
part of the process by which the fashion industry collectively decides what will
be the new trends and the next top models.

Social media and mobile image-sharing platforms, Instagram in particular,
are revolutionizing the fashion industry worldwide, as interest toward new
trends, designers, and products increasingly unfolds online~\cite{Kietzmann2011,
Cassidy2013, Kim2012}. This has obvious implications on the job of fashion
models too. Traditionally, models were not meant to interact directly
with their customers~\cite{mears2005not}. It is instead now customary
for spectators to use Instagram to upload photos or videos during runways
events. This, in turn, has been argued to influence the way fashion designers
design, shoot, and showcase their runways, especially for the case of luxury
brands~\cite{nytAgeOfInstagram}. Famous designers such as
Tommy Hilfiger and Kenneth Cole have been reported to take advantage of
Instagram for customer engagement~\cite{hypebeastInstagram}. 

\rev{As social media become for fashion models a far more important showcase
than magazines and billboards, we wonder if popularity on such platforms can be
used as a proxy to predict success, and seek to answer these research
questions:}

\textsc{rq 1}. \emph{Given data about measurable physical and professional
characteristics of a models, can we predict whether she will be casted for the
upcoming fashion season? }

\textsc{rq 2}. \emph{Does the addition of relevant signals of social media
activity improve the predictability of success of models?}

We tackle these questions using a quantitative approach, with a mix of
exploratory statistics and machine learning experiments. To rule out possible
explanations in terms of cumulative advantage \cite{Rijt2014}, we focus on data
about a group of newcomers, who just started their career in the fashion world. As
a simple and reasonable measure of success, we employ the number of catwalks a
model walked. 

The rest of the paper is organized as follows. We give a brief overview of the
work related to predicting success in cultural markets, and fashion in
particular, in the next section. We then describe the two datasets used in the
study: the ``new faces'' section of the Fashion Model Directory (FMD)
website\footnote{\url{http://www.fashionmodeldirectory.com}} and
Instagram.\footnote{\url{http://instagram.com}} We then present the main
results of this study. We start with a descriptive analysis of the FMD data, and
estimate the degree of association between the tenure of a model and a number of
standard industry metrics using a regression framework. Finally, we describe the
machine learning approach we employed, and the evaluation metrics used to assess
the quality of the predictions of our statistical framework. The paper
concludes describing the broader relevance of our findings to the emerging
``Science of Success'' research field, and potential future directions.

\section{Related Work}

In this study we are looking at predicting popularity of fashion models using
data from social media activity. The study of trends in cultural markets has a
long-standing tradition~\cite{Bikhchandani1992}, and various researchers have
exploited social data from a wide range of backgrounds. For example, Asur and
Huberman~\cite{asur2010predicting, asur2011trends} used Twitter data to predict
the box-office performance of newly-released blockbuster movies, and later
Mesty\'an et al.~\cite{mestyan2013early} improved these results using data from
Wikipedia. Ferrara et al.~\cite{ferrara2013clustering, ferrara2013traveling,
jafariasbagh2014clustering} studied the emergence of information trends in
social media settings, and the rise to popularity of Instagram users in
photography contexts~\cite{ferrara2014online}. 

In contrast, in the context of fashion, it is worth noting that practical
application of trend-detection technologies has started to become possible only
in recent years, thanks to the increasing availability of online user-generated
data about fashion apparel trends~\cite{lin2014hidden, lin2015styles,
lin2015text, Green2015}. 

Instagram was the platform selected in our study. It is a mobile image-sharing
service that specializes in instant communication of trends and visual
information in general~\cite{ferrara2014online}, in a way similar to Twitter,
Pinterest, and Flickr. Besides the detection of trends, a wide range of aspects
related to image sharing have received attention from the research community,
such as depression and online behavior~\cite{Andalibi2015,Wang2015}, situated
usage in museums~\cite{Hillman2015,Weilenmann2013}, or organization of
information through tags~\cite{Nov2008,Klemperer2012,Gilbert2013}. 

\section{Methods}
\subsection{Fashion Model Directory}

\begin{figure}[!t]\centering
    \includegraphics[width=.9\columnwidth]{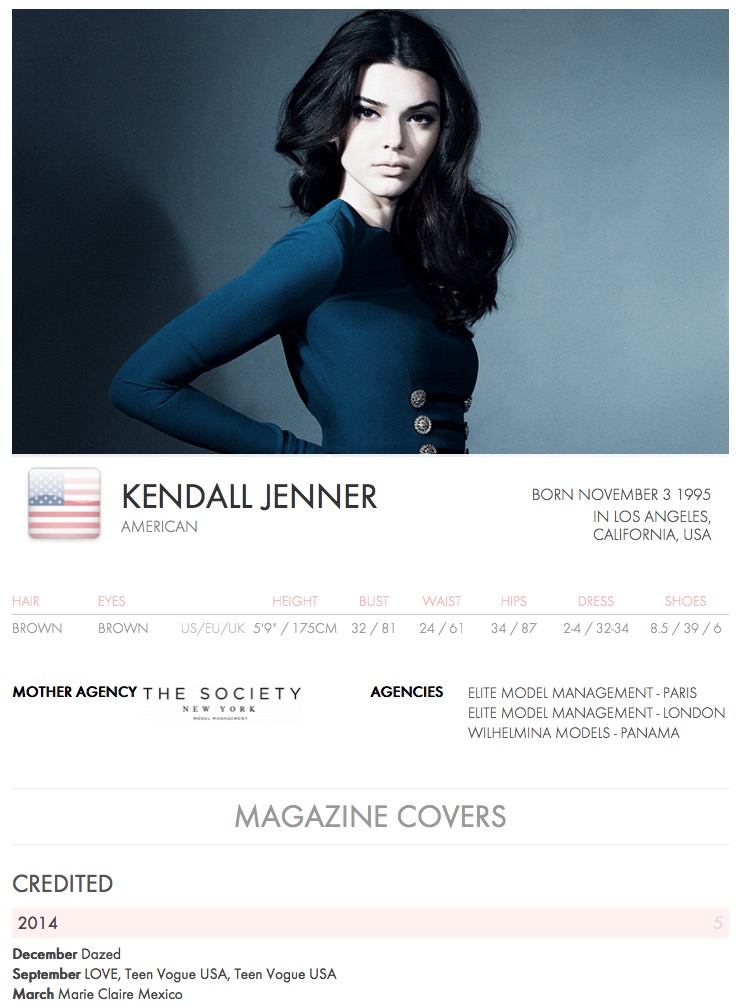}
    \caption{Example of a FMD profile page. Image \textcopyright~FMD -- The Fashion Model Directory} 
    \label{fig:fmd_page_ex}
\end{figure}

We collected data from the Fashion Model Directory (FMD) website, one of the
largest fashion databases of professional female fashion models~\cite{wiki:fmd}.
Figure~\ref{fig:fmd_page_ex} shows the profile of one of the fashion models on
FMD as an example. As can be seen from the figure, FMD profiles, similarly to a
r\'esum\'e, provide a mix of biographical, physical, and professional
experience information, notably casting agencies and walked runways. 

While the database claims to include profiles for over 10,000 models, our
analysis focuses only on its recent additions, which are listed under the
category ``New Faces''. We collected our dataset in December 2014, finding $N =
431$ new faces for the 2015 Spring\,/\,Summer (S\,/\,S) season.

For each model the FMD data thus consists of the following attributes: name,
hair color, eye color, height, hip size, dress size, waist size, shoes size,
list of agencies, nationality, and details about all runways the model walked on
(year, season, and city). We discarded the data about hair and eye color, as
the color coding was not reliable enough to allow for a meaningful
characterization of these features. For similar reasons, we also discarded
nationality information. All body sizes (height, hips, dress, waist, shoes) were
converted in the metric system. For shoe size we used the Paris point units
system. Furthermore, the data were cleaned, substituting, in two cases where the
data about shoe and waist size were missing, the missing values with the group
average. Before running regressions, any non-categorical variable was also
centered around the mean and standardized. 

To account only for a homogeneous set of runway experiences, we considered only
runways occurred during the fashion weeks for the 2015 S\,/\,S season in New
York, London, Paris, and Milan. These occurred during the period of September
2014 and were the most recent major fashion weeks at the time of our data
collection (December 2014). We found that, collectively, the new faces in our
sample had performed $W = 1402$ runway walks (3.25 walks per model on average)
for $B = 313$ distinct branded runway events.

Finally, we annotated each agency to reflect its reputation in the fashion
industry. We use a simple binary classification system, by which high-prestige
agencies are assigned to the ``top agency'' category. Since FMD does not provide
this information, we retrieved a list of all top agencies from another online
fashion database, Models.com,\footnote{\url{http://www.models.com}.} which
collects experts knowledge to determine the reputation and influence of casting
agencies. As a result, 329 out of 431 models were hired by at a least one top
agency, 87 models were not hired by any top agency, and 15 had no agency
information in their profile (which could indicate that either the model was not
using any agency, or that the information was simply missing from the database;
see Table \ref{tab:instagram_account}).

\begin{figure}[t]
    \centering
    \includegraphics[width=\columnwidth]{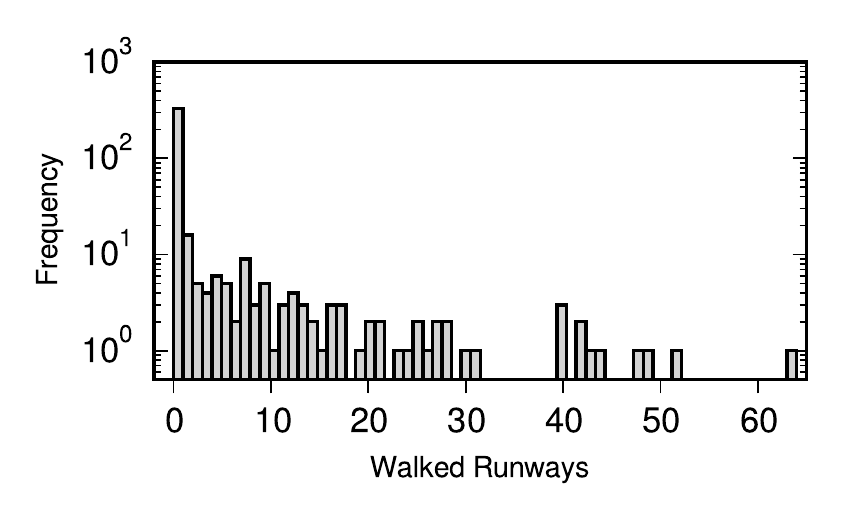}
    \caption{Distribution of walked runways by new faces of the S\,/\,S prior to
    the Fashion Week season.}
    \label{fig:runway_dist}
\end{figure}

The majority of new models did not perform even a single runway during the four
fashion weeks in September 2014; only around 24\% of models (102 out of 431)
performed at least one runway, with a small minority having participated to
several runways (see Figure~\ref{fig:runway_dist}). This is not surprising, and
further suggests that there exists a strong popularity bias in castings, in line
with previous work~\cite{godart2009cultural}. 

\newpage\subsection{Instagram}

We collected data about the social media presence of our FMD new faces using
Instagram. We found $N_i = 253$ Instagram accounts (59\%), accounting for
$W_i=1181$ runway walks (4.67 walks per model on average). Models under contract
with a top agency are more represented on Instagram (65\%) than those with a
non-top agency (41\%), and no agency at all (40\%), see
Table~\ref{tab:instagram_account}. 

\begin{table}[tb]
  \centering\small
  \begin{tabular}{@{}l@{} c c c r@{}}
    \toprule
    & \tabhead{Top agency} & \tabhead{Non-top agency} & \tabhead{No agency} &
    \tabhead{Total} \\
    \midrule
    with Instagram & 214 & 33 & 6 & 253\\
    w/o Instagram & 115 & 54 & 9 & 178 \\
    \midrule
    Total & 329 & 87 & 15 & 431\\
    \bottomrule
  \end{tabular}
  \caption{New faces of the 2015 S\,/\,S season under contract with at
  least one top agency and their presence on Instagram.} 
  \label{tab:instagram_account}
\end{table}

Using the media endpoint of the Instagram API, we collected all media posted by
any FMD new face in the three-month period before September 4th, 2014, the
beginning of the New York Fashion Week. Metadata of each posted media include
the number of likes and comments, as well as the the metadata of the first 125
likes of each post (\emph{e.g.}, time stamp of the like, name of the liking user,
etc.). We then computed the minimum, maximum, median, and average number of
likes and comments of all posts uploaded by each new face, as well as the number
of posts during the period, for the three months before and after the fashion
week events. Similarly to the FMD data, all variables were standardized before
using them in regressions.

\subsection{Sentiment Analysis}

We supplement the analysis of social media activity with sentiment
analysis. To do so, we selected only comments written in English. Language was
detected using a simple Naive Bayes classifier~\cite{nakatani2010langdetect}.
\rev{We extracted the comments on posts uploaded before the Fashion Week season, and
calculated the average sentiment score of each model using \emph{Vader}, a
state-of-the-art, rule-based algorithm~\cite{hutto2014vader}. We included only
those models who received at least one comment written in English to any of
their posts, finding a subset of $N_s = 198$ models, who account for $W_s =
1052$ runway walks (5.31 walks per model on average). \emph{Vader} is designed
to deal with social media data, as it is based on a manually-defined vocabulary
that encodes grammatical and syntactical conventions common to online documents.
It is capable of capturing sentiment intensity with an accuracy of 84\%, which
outperforms other algorithms as well as individual human raters.}

\subsection{Predictive classification of success}

To forecast success within the new faces cohort, we performed a binary
classification exercise. Since most models in the cohort did not walk any
runway, to avoid further class imbalance we consider two classes: models with
zero walks (unpopular) and models with one or more walks (popular). To learn the
predictive score distributions, we applied three widely-used machine learning
algorithms, based on ensemble methods and boosting: Decision Tree (DT)
(baseline), Random Forest (RF) \cite{breiman2001random}, and AdaBoost (AB)
\cite{freund1995desicion}. We used the implementations of
scikit-learn~\cite{pedregosa2011scikit}, optimally tuning the parameters as
follows: entropy is used to measure the quality of the decision splits; all
statistical models employ 25 estimators; DT and RF have a pruning setting of max-depth to 5,
and RF adopts a maximum of 5 features. This framework allows us to evaluate the
forecasting power of the various sets of features presented above using standard
performance metrics, such as AUROC (Area Under the Receiver Operating
Characteristic curve), and accuracy scores. We report results obtained by
averaging one thousands iterations of $k$-Fold Cross Validation ($k=5$) in which
80\% of data is used to train the statistical models and the remainder 20\% is used for
prediction. The last experiment of this paper, however, represents a ``real''
prediction task in which we train the machine learning models with the previous
fashion season (2015 S\,/\,S) data, and use them to predict the upcoming 2015-16
Fall\,/\,Winter (F\,/\,W) season.

\section{Results}

\subsection{Descriptive analysis}

\begin{figure*}[t!]
    \centering
    \includegraphics[width=.95\textwidth,clip=true,trim=0 15 20 15]{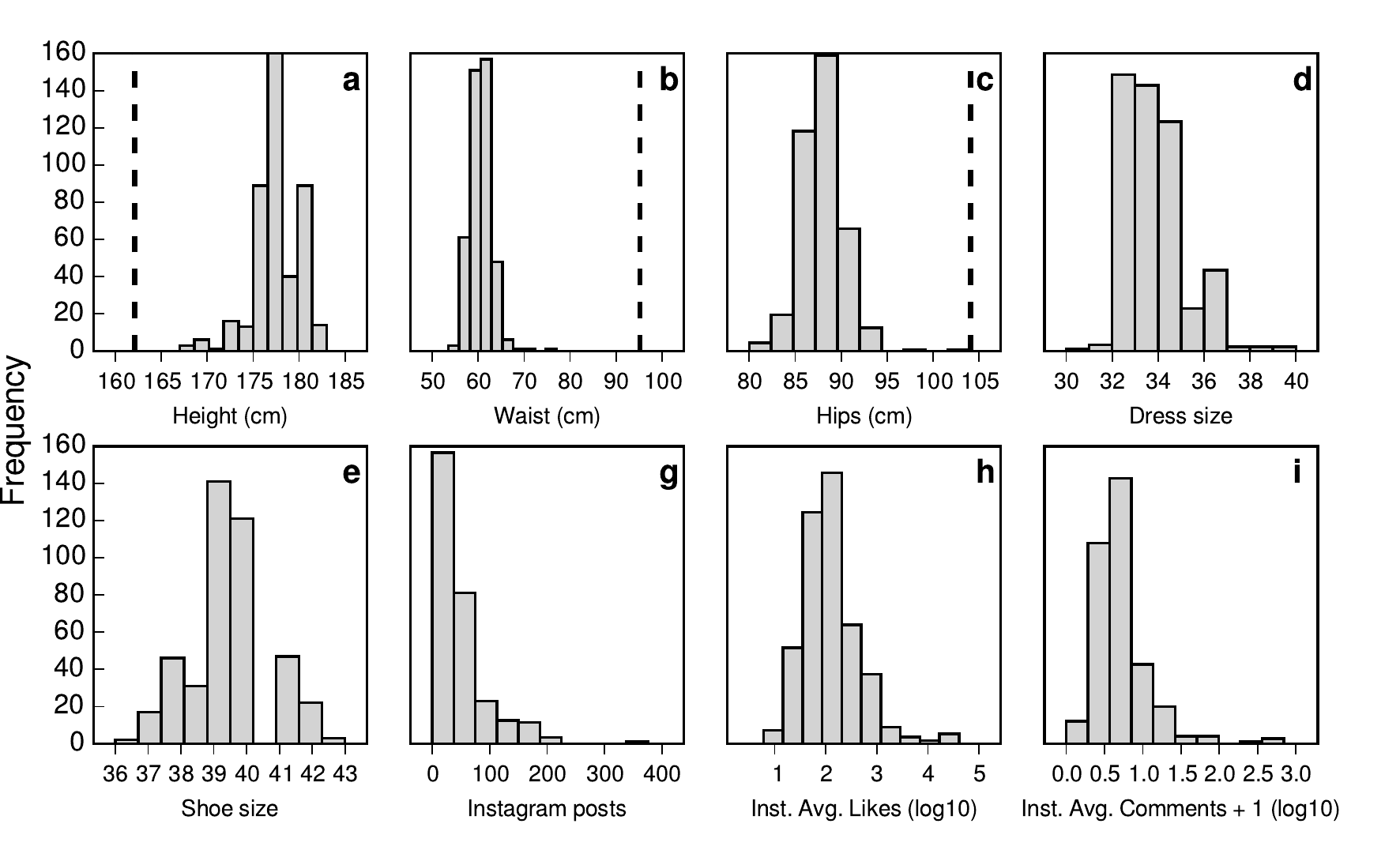}
    \caption{Distribution of body size measures and Instagram activity of new
        faces of the 2015 S\,/\,S season ($N=431$). Dashed lines, where shown,
        indicate averages computed on the closest matching age group in the US female
    population.}
    \label{fig:distributions}
\end{figure*}

We first analyzed height, hips, dress, waist, and shoe size of the new faces
group and assessed whether there is any obvious association with the number of
walked runways. Figures~\ref{fig:distributions}(a)--(e) show the distribution of
body size measurements and, where available
(Figures~\ref{fig:distributions}(a)--(c)) how these features compare to the US
female population for the closest age group (20--29 y.o. for height and waist
size~\cite{fryar2012anthropometric}; 18--25 y.o. for hip
size~\cite{nytHipSize}). Here the data have been plotted before rescaling. The
variables are all distributed within narrow ranges, following approximately
normal distributions. Unsurprisingly, even the shortest model in our dataset is
much taller than the US female average. \rev{In general, as far as body measures are
concerned, our group of new faces seems to represent a very biased sample of the
general female population}. Table~\ref{tab:phy_description} reports standard
descriptive statistics of the sample.

\begin{table}[tb]
  \centering
  \small
  \begin{tabular}{@{} l r r r r r @{}}
    \toprule
    & \tabhead{Height (cm)} & \tabhead{Hips (cm)} & \tabhead{Dress} & \tabhead{Waist (cm)} & \tabhead{Shoes}  \\
    \midrule
    Mean    & 177.48 & 87.98  & 33.36 & 60.49 & 39.44 \\
    Std. Dev& 2.49   & 2.33   & 1.36  & 2.17  & 1.19 \\
    Min.    & 167.00 & 80.00  & 30.00 & 53.50 & 36.00 \\
    Median  & 178.00 & 88.00  & 33.00 & 60.00 & 39.00 \\
    Max.    & 183.00 & 104.00 & 40.00 & 77.00 & 43.00 \\
    \bottomrule
  \end{tabular}
  \caption{Body size measures of new faces of the 2015 S\,/\,S season
  ($N=431$).}
  \label{tab:phy_description}
\end{table}

We study the association between body size measures and the main dependent
variable, the number of runways the models walked \emph{prior} to the 2015
S\,/\,S fashion week season, using a regression framework. 

Since our measure of success is based on the count of walked runways, we used a
Poisson regression model to estimate the chances of walking a single runway as a
function of various regression features. We start by only considering physical
attributes as regressors. The results are reported in Table~\ref{tab:reg_result}
(see Model 1). The expected number of walked runways for the baseline new face
is 2.25. Height is positively associated to increased chances of walking a
runway --- specifically, 2.27 times for approximately each additional cm,
relative to the group average baseline. Larger dress, hips, and shoe sizes are
all negatively associated with the chances of walking a runway, while waist size
seem to be not associated in either way. 

In Table~\ref{tab:phy_corr} we also report correlations between all body size
regressors, as a simple test for possible source of multicollinearity. We find
that dress, waist, and hip size are pairwise correlated with each other, as
well as shoe size with height. All correlations appear to be of moderate entity.

\begin{table}[tb]
  \centering
  \small
  \begin{tabular}{@{} l r r r r r @{}}
    \toprule
    & \tabhead{Height} & \tabhead{Hips}  & \tabhead{Dress} & \tabhead{Waist} & \tabhead{Shoes} \\
    \midrule
    \tabhead{Height} & $1.00$  & & & & \\
    \tabhead{Hips}   & $0.01$  & $1.00$  & & & \\
    \tabhead{Dress}  & $-0.06$ & $0.25$  & $1.00$ & & \\
    \tabhead{Waist}  & $0.02$  & $0.58$  & $0.28$ & $1.00$ & \\
    \tabhead{Shoes}  & $0.36$  & $-0.01$ & $0.00$ & $0.07$ & $1.00$ \\
    \bottomrule
  \end{tabular}
  \caption{Pairwise correlations between body size measures of the new faces of
      the 2015 S\,/\,S season ($N=431$)}
  \label{tab:phy_corr}
\end{table}

Adding the information on agencies (see Model 2), we find a strong association
between having a top agency and the number of walked runways: models with a top
agency have, everything else being equal, nearly ten times higher chances ($\exp\left(
 2.29\right) = 9.87$) of walking a runway, than their counterparts represented
 by non-top agencies. The chances of walking for models without a prestigious
 agency drop substantially, as the expected count for the baseline is now only
 0.28 walks. This is consistent with previous
 research~\cite{godart2009cultural}, and highlights the role of agencies in
 setting fashion models trends in the fashion industry.

We then focus on models with an Instagram account ($N_i=253$) and assess how the
average number of posted media, received likes, and comments is associated to
the count of walked runways. Figures~\ref{fig:distributions}(g)--(i) show
distribution histograms (log10-scaled) of these variables.

Model 3 and Model 4 replicate the above findings on the subset of new faces with
an Instagram account. In particular, within this subset the gap between those
with a top agency and those without is even more marked. Adding the
Instagram-related variables does not change much the association with height,
hip size, and agency. Instagram activity seem to have mixed associations with
runway walks. Additional posts over the average activity yield a 15\% higher
chances of walking a runway but, surprisingly, more likes tend to \emph{lower}
the chances of walking a runway (about 10\% less). The average number of
Instagram Likes is highly correlated with the average number of comments ($r =
0.82$) yet it has a negligible correlation with the number of posts ($r =
0.15$). The number of received comments does not seem to be correlated with the
number of posts ($r = 0.00$).

The overall picture does not change significantly when we look at the sentiment
expressed by Instagram users when commenting on the media posted by our new
faces (Model 6, 7, and 8 of Table~\ref{tab:reg_result}). The sentiment itself
appears to be positively associated to better chances of walking a runway (23\%
more), together with the overall number of comments.

Including the information about agency and all Instagram-related variables
yields better statistical models for both the overall sample of new faces and
those with an Instagram account, as shown by both the Akaike (AIC) and Bayesian
Information criterion (BIC) scores. This indicates that all variables
potentially provide useful predictive signals. In the next section we describe
the results of the prediction tasks in detail.

\begin{table*}[tb]
\centering
\newcolumntype{d}[1]{D{.}{.}{#1} }
\small
\begin{tabular}{@{} l @{\extracolsep{4em}} d{2} @{\extracolsep{\fill}} d{2} @{\extracolsep{4em}} d{2} @{\extracolsep{\fill}} d{2} d{2} @{\extracolsep{4em}} d{2} @{\extracolsep{\fill}} d{2} d{2}}
\toprule
& \multicolumn{2}{c}{\tabhead{All new faces}}
& \multicolumn{3}{c}{\tabhead{w/ Instagram}}
& \multicolumn{3}{c}{\tabhead{w/ Instagram \& Sentiment}}\\
& \multicolumn{2}{c}{\tabhead{($N=431, W=1402$)}}
& \multicolumn{3}{c}{\tabhead{($N_i=253, W_i=1181$)}}
& \multicolumn{3}{c}{\tabhead{($N_s=198, W_s=1052$)}}\\
\midrule
& \multicolumn{1}{c}{Model 1} 
& \multicolumn{1}{c}{Model 2}
& \multicolumn{1}{c}{Model 3}
& \multicolumn{1}{c}{Model 4}
& \multicolumn{1}{c}{Model 5}
& \multicolumn{1}{c}{Model 6}
& \multicolumn{1}{c}{Model 7}
& \multicolumn{1}{c}{Model 8}\\
\midrule
Intercept       &  0.81\sss & -1.27\sss &  1.21\sss & -7.70\s   & -7.70\s   &  1.32\sss & -7.59     & -7.65\s    \\
Height	        &  0.82\sss &  0.76\sss &  0.89\sss &  0.85\sss &  0.88\sss &  0.89\sss &  0.86\sss &  0.90\sss  \\
Dress           & -0.23\sss & -0.17\sss & -0.13\sss & -0.05     & -0.05	   & -0.16\sss & -0.13\sss & -0.12\sss  \\
Hips            & -0.35\sss & -0.33\sss & -0.24\sss & -0.28\sss & -0.29\sss & -0.23\sss & -0.33\sss & -0.36\sss  \\
Waist           & -0.04     & -0.00     &  0.06\s   &  0.11\sss &  0.10\sss &  0.10\sss &  0.18\sss &  0.19\sss  \\
Shoes           & -0.38\sss & -0.37\sss & -0.35\sss & -0.37\sss & -0.36\sss & -0.41\sss & -0.44\sss & -0.45\sss  \\ 
Has Top Agency  &           &  2.29\sss &           &  9.05\ss  &  9.04\ss  &           &  9.03\s   &  9.07\ss   \\
Inst. Posts     &           &           &           &           &  0.14\sss &           &           &  0.08\ss   \\
Inst. Likes     &           &	        &           &           & -0.18\sss &           &           & -0.17\sss  \\
Inst. Comments  &           &           &           &           &  0.24\sss &           &           &  0.21\sss  \\
Inst. Sentiment &           &           &           &           &           &           &           &  0.16\sss  \\
\midrule
AIC & 4814.41 & 4531.28 & 3339.79 & 3064.96 & 3038.96 & 2734.13 & 2489.47 & 2462.77 \\
BIC & 1824.81 & 1545.74 & 1639.85 & 1368.52 & 1353.12 & 1426.88 & 1185.46 & 1171.92	\\
\bottomrule
\end{tabular}
\caption{Poisson regression results for the new faces of the 2015 S\,/\,S season.
    Dependent variable is the count of runways walked. Legend: ${}^{*}:p<0.05$; ${}^{**}:p<0.01$;
${}^{***}:p<0.001$.}
  \label{tab:reg_result}
\end{table*}

\subsection{Forecasting success in fashion}

For each classification algorithm (DT, RF, and AB) we learned three distinct
predictive models: \emph{(i)} with only body size measures (height, hips, dress,
waist, and shoes) (\textsc{body}); \emph{(ii)} with physical attributes and the
binary information about whether the fashion model has a top agencies or not
(\textsc{body+agency}); and, \emph{(iii)} with body size measures, agency
information, and Instagram-related signals \rev{---number of posts, average number
of likes and comments received--- 
(\textsc{body+agency+insta})}. For the latter statistical model, we restrict the
training data to use only the media posted in the three months \emph{before} the
fashion week. As shown in Table~\ref{tab:mlclf_result}, and consistently with
results from the previous section, when trained on 2015 S\,/\,S runway walks
data, social media features improve accuracy of the statistical model. According
to $t$-tests, all improvements are statistically significant.
\rev{We also tried other statistical models~\cite{pedregosa2011scikit} (SVM,
Logistic Regression, Naive Bayes, etc.): none yielded AUROCs or accuracy above
60\%.}

\begin{table*}[tb]
    \centering
    \small
    \begin{tabular}{l*{2}{l} *{4}{@{\extracolsep{2em}}r @{\extracolsep{\fill}}l}}
        \toprule
        & \multicolumn{2}{c}{\tabhead{\textsc{body}}}
        & \multicolumn{4}{c}{\tabhead{\textsc{body+agency}}}
        & \multicolumn{4}{c}{\tabhead{\textsc{body+agency+insta}}}\\
& \textsc{acc}	
& \textsc{roc}	
& \multicolumn{2}{c}{\textsc{acc}}
& \multicolumn{2}{c}{\textsc{roc}}	
& \multicolumn{2}{c}{\textsc{acc}}
& \multicolumn{2}{c}{\textsc{roc}} \\ 			   
    \midrule
    Decision Tree & 0.596 & 0.558 & 0.635 & (+0.039)\sss & 0.563 & (+0.005)\sss  & 0.694 & (+0.059)\sss   & 0.619 & (+0.056)\sss \\
    Random Forest & 0.643 & 0.549 & 0.656 & (+0.013)\sss & 0.586 & (+0.037)\sss & 0.733 & (+0.077)\sss & 0.688 & (+0.102)\sss \\
    Ada Boost     & 0.640 & 0.533 & 0.636 & (-0.004)\sss & 0.556 & (+0.023)\sss  & 0.692 & (+0.056)\sss & 0.640 & (+0.084)\sss \\
    \bottomrule
    \end{tabular}
    \caption{\rev{Accuracy (\textsc{acc}) and Area Under the \textsc{roc} curve
    (\textsc{roc}) values for all classifiers.} Increments for the classifier
    with all features (\textsc{body+agency+insta}) are computed over that
    without Instagram-related features (\textsc{body+agency}), which is in turn
    computed over the baseline (\textsc{body}). All improvements are
    statistically significant. \rev{Random Forest is the model with the best
    predictive power, scoring a top accuracy of 73.3\% and an AUROC of 68.8\%.}
    Legend: ${}^{*}:p<0.05$; ${}^{**}:p<0.01$;
    ${}^{***}:p<0.001$.}
    \label{tab:mlclf_result}
\end{table*}

To test the actual forecasting power of our framework based on the classifiers
trained on the 2015 S\,/\,S data, we attempt to predict the popularity labels
for the next season, the 2015-16 F\,/\,W Fashion Week, that is, on a completely
separate test set. To do so, \rev{we manually collected a new and more recent
dataset (May 2015) containing the \emph{new faces} of the latest fashion season,
and up-to-date information about runways performed during the 2015-16 F\,/\,W
Fashion Week} (February 12--March 11, 2015). We found 15 such new face profiles.
This set is roughly balanced (8 fashion models ran at least one top walk, 7 did
not appear in any of the four main events), and each profile links to an
Instagram account, allowing us to employ all predictive features
(\textsc{body+agency+insta}). 

\begin{figure}[t!]
    \centering
    \includegraphics[clip=true,trim=45 0 0 0]{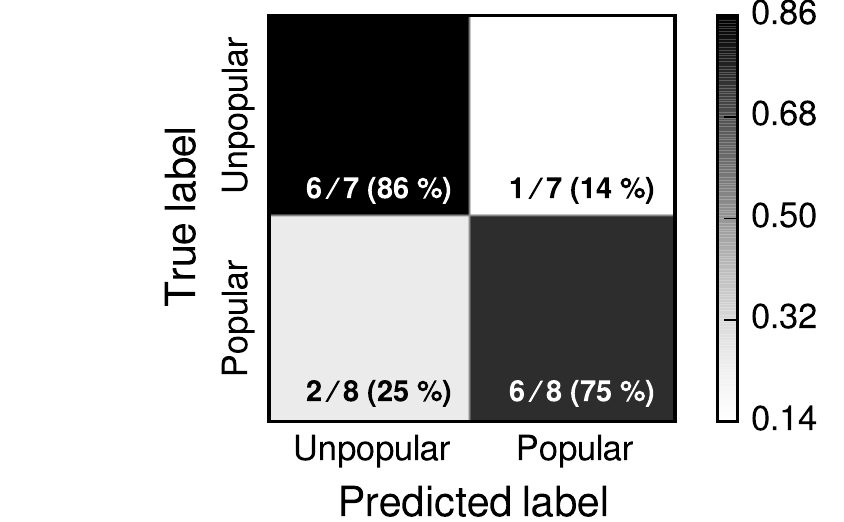}
    \caption{\rev{Confusion matrix of the Random Forest performance for the
    prediction task (2015-16 F\,/\,W Fashion Week).}} \label{fig:cm}
\end{figure}

Social media features for the validation test set were built using the
meta-data of media posted in the three months before the season only (November 12,
2014 to February 11, 2015). \rev{The results for the best predictive model,
Random Forest (RF), along with the true popularity labels (\emph{became
popular}), are shown in Table~\ref{tab:test_result}. Random Forest scores an
AUROC performance above 81\%: impressively, RF is able to correctly predict 6
out of 8 fashion models who became popular during the 2015-16 F\,/\,W, using
training data from the past season only. Random Forest also successfully
identified 6 of the 7 fashion models who did not perform in any top event. The
confusion matrix in Fig.~\ref{fig:cm} summarizes these results.}

\begin{table*}[tb]
\centering
\small
\begin{tabular}{@{}cccccccccc|c|c@{}}
\toprule
  \tabhead{Fashion}	
& \tabhead{Height} 
& \tabhead{Hips} 
& \tabhead{Waist} 
& \tabhead{Dress} 
& \tabhead{Shoes} 
& \tabhead{Instagram} 
& \tabhead{Instagram} 
& \tabhead{Instagram} 
& \tabhead{Has top}
& \tabhead{Became} 
& \tabhead{RF Prediction} \\
  \tabhead{Model ID}
&
&   
&   
&   
&   
& \tabhead{Posts} 
& \tabhead{Comments} 
& \tabhead{Likes} 
& \tabhead{agency} 
& \tabhead{popular}  
& \\
\midrule
    1  & 178 & 86.5 & 58 & 33 & 41.0 &  59 &  4 & 148 & \T & \T & \T \\
    2  & 178 & 86.0 & 60 & 33 & 40.0 &  24 &  0 &  32 & \F & \T & \F \\
    3  & 179 & 88.0 & 61 & 34 & 39.0 &   0 &  0 &   0 & \T & \T & \F \\
    4  & 180 & 89.0 & 60 & 34 & 41.0 &  52 &  1 &  93 & \T & \T & \T \\
    5  & 175 & 86.0 & 58 & 33 & 38.0 & 163 &  2 &  70 & \T & \F & \F \\
    6  & 180 & 89.0 & 60 & 34 & 41.0 &   2 &  7 &  48 & \T & \T & \T \\
    7  & 180 & 90.0 & 61 & 34 & 40.0 &  10 &  3 & 116 & \T & \T & \T \\
    8  & 178 & 87.0 & 61 & 33 & 39.0 &  34 &  2 &  90 & \T & \T & \T \\
    9  & 183 & 86.0 & 62 & 34 & 41.0 &  16 &  2 &  61 & \T & \T & \T \\
    10 & 176 & 87.0 & 59 & 33 & 38.5 &  17 & 17 & 647 & \T & \F & \T \\
    11 & 177 & 86.0 & 60 & 32 & 38.5 &  38 &  2 &  51 & \T & \F & \F \\
    12 & 180 & 90.0 & 60 & 33 & 40.0 &  29 &  1 &  59 & \F & \F & \F \\
    13 & 169 & 88.0 & 60 & 33 & 38.0 &  49 &  9 & 570 & \T & \F & \F \\
    14 & 179 & 94.0 & 65 & 35 & 41.0 &  58 &  3 &  52 & \F & \F & \F \\
    15 & 180 & 83.0 & 62 & 35 & 43.0 &  11 & 15 & 546 & \F & \F & \F \\
\bottomrule
\end{tabular}
  \caption{\rev{Performance of our predictive models trained on 2015 S\,/\,S
  data, and tested on 15 new fashion models who appeared in the 2015-16 F\,/\,W
  Fashion Week. Our best classifier, Random Forest, correctly predicts 6 out of
  8 positive instances (became popular), and 6 out of 7  negative ones (not
  becoming popular) yielding 80\% accuracy and an AUROC score of 81.25\%.}}
  \label{tab:test_result}
\end{table*}

\newpage
\rev{
\subsection{The elements of success in fashion modeling }

\begin{figure*}[t]
    \centering
    \subfloat[Fashion Model  1\label{modelPic:1}]{%
        \includegraphics[width=0.5\columnwidth]{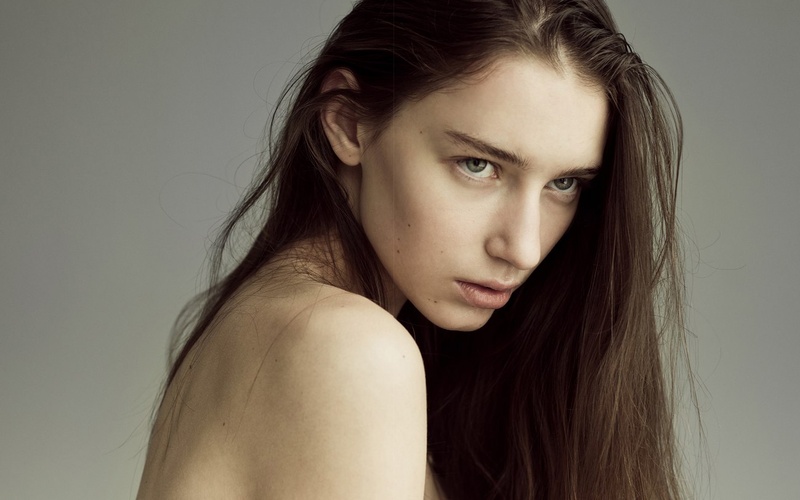}
    }
    \subfloat[Fashion Model  4\label{modelPic:4}]{%
        \includegraphics[width=0.5\columnwidth]{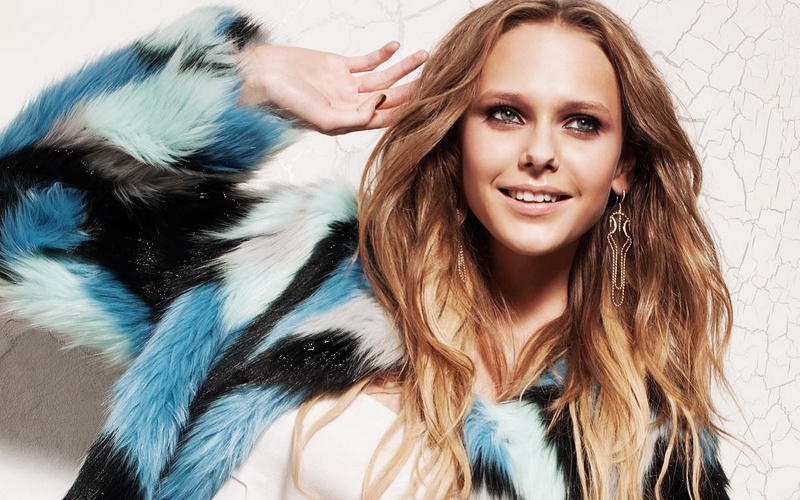}
    }
    \subfloat[Fashion Model  6\label{modelPic:6}]{%
        \includegraphics[width=0.5\columnwidth]{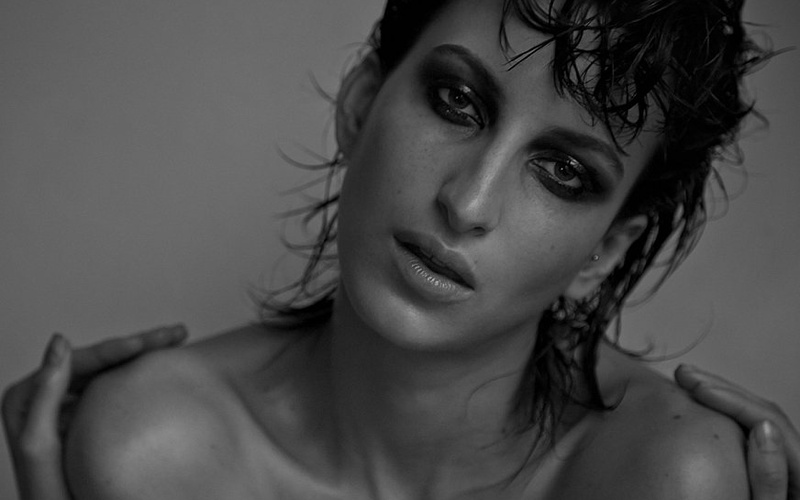}
    }\\
    \subfloat[Fashion Model  7\label{modelPic:7}]{%
        \includegraphics[width=0.5\columnwidth]{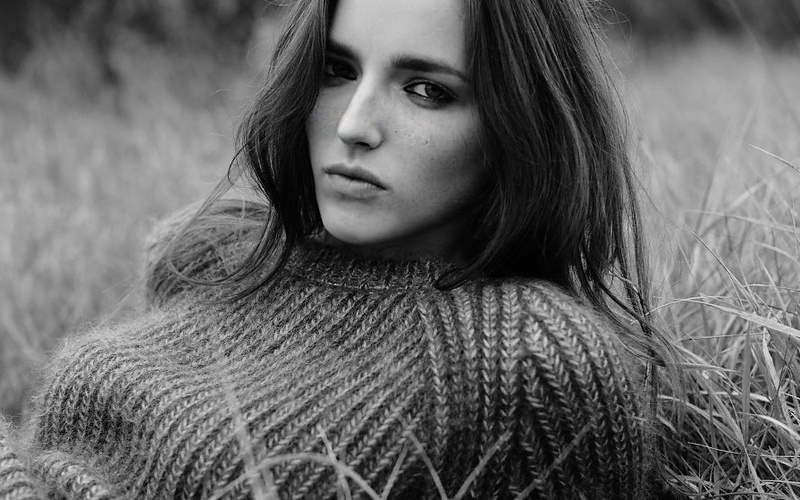}
    }
    \subfloat[Fashion Model  8\label{modelPic:8}]{%
        \includegraphics[width=0.5\columnwidth]{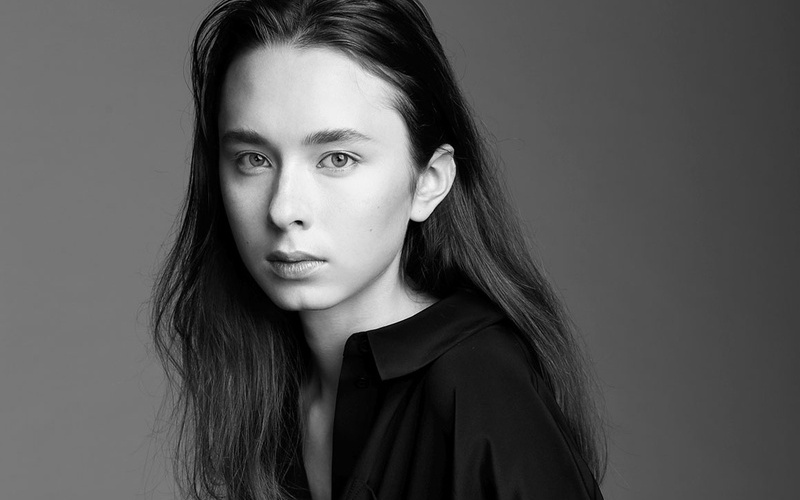}
    }    
    \subfloat[Fashion Model  9\label{modelPic:9}]{%
        \includegraphics[width=0.5\columnwidth]{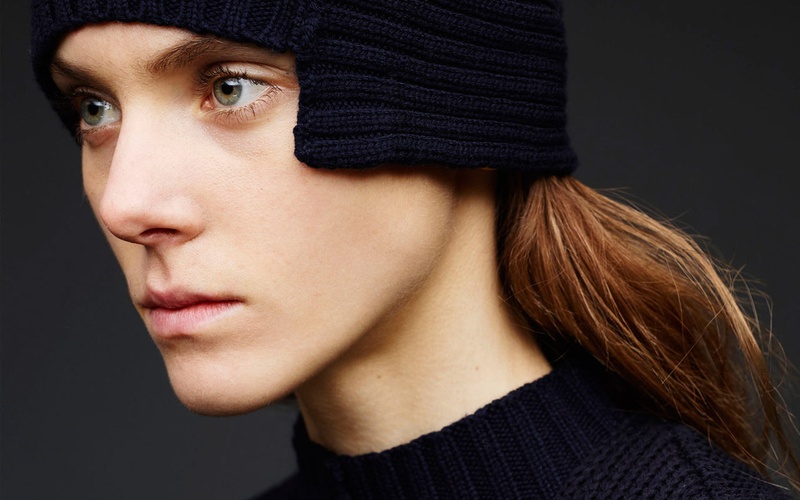}
    }    
    \caption{\rev{FMD profiles of the six \emph{new faces} whose success
(having at least one runway during 2015-16 F\,/\,W Fashion Week) was correctly
predicted by our framework.} All images \textcopyright~FMD -- The Fashion Model
Directory.} 
\label{fig:modelImgs}
\end{figure*}

The analysis of the prediction task provides some interesting insights: Fashion Models 
11 and 13, although having top agencies, did not rise to popularity; Fashion Model 5,
however, became popular even without a top agency: both of these dynamics have
been correctly captured by our prediction framework. It is worth considering
when our framework failed. Fashion Models  2 and 3 represent the two \emph{false
negatives}: they both exhibit low social media activity, a signal highly
regarded by our predictor (see below), which induces Random Forest to mistake.
Fashion Model 10, the only \emph{false positive}, on the other hand shows very high
social media engagement levels, yet did not perform in any top runway during the
2015-16 F\,/\,W season. The FMD profile pictures of the six models whose success
was correctly predicted by our framework are shown in
Figure~\ref{fig:modelImgs}. 
} 

To understand which features contribute most to the predictive signal we compute
feature importance. The contribution of each feature is here calculated as
information gain. The results for Random Forest (RF) are shown
in Table~\ref{tab:feature_importance}. \rev{For prediction purposes, we note how
the three social media activity features contribute as much as having optimal
physical attributes and more than being under contract with a top agency. The
top 3 features used by RF are: \emph{(i) Instagram \#Likes}, \emph{(ii) Height},
and \emph{(iii) \#Posts}. Similar results hold for other classifiers.}

\begin{table}[b]
    \centering
    \small
    \begin{tabular}{lcc}
        \toprule
        \tabhead{Feature} &
        \tabhead{Type} &
        \tabhead{Importance} \\
        \midrule
        Height & Physical & 0.16\\
        Dress & Physical & 0.05\\
        Hips & Physical & 0.09\\
        Waist & Physical & 0.10\\
        Shoes & Physical & 0.09\\
        Has Top Agency & Professional & 0.05\\
        Instagram Posts & Social & 0.16\\
        Instagram Comments & Social & 0.13\\
        Instagram Likes & Social & 0.18\\
        \bottomrule
    \end{tabular}
    \caption{\rev{Feature importance (Random Forest model) to predict fashion models' success.}}
    \label{tab:feature_importance}
\end{table}

\section{Discussion}

Social media are increasingly used as sensors of social collective
phenomena~\cite{asur2010predicting,ferrara2013traveling}. Increasingly, the
usage of social data, often in conjunction with other data sources, proves
crucial to be able to represent real-world events, trends, information
diffusion, and social behavior. In this study we were concerned with understanding
whether it is possible to predict fashion models popularity, complementing
physical and professional information with social data. 

Our methodology has of course limitations, and we here report few notable ones:
\begin{itemize}
    \item All brands during the season are equally treated. Runways of higher
        reputation brands, such as Herm\`es or Chanel, should be reflected with
        higher weights if compared to new and relatively unpopular brands. We
        plan to incorporate such prestige in future revisions of our statistical
        models, and observe what effects this yields.
    \item Our measure of popularity only takes into account the number of
        runways walked. This neglects several aspects of popularity within the
        fashion industry, such as appearances on magazines and social events. We
        plan to incorporate further dimensions of success in future work, to
        determine how these additional dimensions play along with the success
        measured by runways.
    \item \rev{Our ``real''  prediction task is tested on a very small dataset
        containing only 15 fashion models appeared during the 2015-16 F\,/\,W
        Fashion Week: although this limitation is due to the intrinsic scale of
        fashion events and to our data sources, more data in the future will be
        needed to determine the general performance of our framework.}
    \item \rev{Our study is confined to one single online platform, Instagram:
        its peculiar characteristics (\emph{e.g.}, the mobile-oriented nature) might
        affect the dynamics of content generation and perceived popularity, as
        opposed to other platforms with different usage purposes, like
        information sharing (Twitter~\cite{kwak2010twitter}) or befriending
        activities (Facebook~\cite{de2014facebook}).}
       
    \item \rev{Finally, our study is limited to analyze only female fashion, while man modeling
    is increasingly becoming more mainstream. It will be interesting to see, when data become 
    available, whether our results apply to the male fashion modeling market as well.}
\end{itemize}

\newpage
\section{Conclusions}

The ingredients to career success oftentimes remain mysterious. In the fashion
industry, style is often credited as that ineffable quality all successful
individuals have. The present contribution shows how a number of seemingly
disconnected characteristics are actually tightly entangled: physical attributes
are required for inclusion in the modeling profession, but do not suffice. The
professional contribution of trend-setting top agencies play an equally
important role. And, as we first show in this paper, in the new era of social
networks, online presence helps succeed, as we see by the improvement in the
predictive power of our forecasting models.

We submit a few possible explanation to this observation: in a world with
limited attention~\cite{Ciampaglia2015a}, information cascades and the wisdom of
the collectives are precious indicators for casting agencies, promoters,
marketeers, agents, recruiters, and the fashion industry in general. The
response of the online audience  plays an increasingly important role in the
offline fashion industry world: a rising star in the online world will hardly be
ignored, and will probably be noticed by a top agency, facts that will enhance
her likelihood to succeed. \rev{In other words, buzz on social media is a proxy
for the buzz in the offline world, and this reduces uncertainty on the part of
the industry.}

Yet, it remains interesting that, in the regression models, increased activity
on social media had only a weak association with heightened success (though on
average fashion models with an Instagram account tended also to have done more
shows). Perhaps, even these small differences have more chances of getting
amplified due to word of mouth and collective attention, so that social media
may be just facilitating the information cascades mentioned before. \rev{Lacking
data on the word of mouth among industry professionals, in this work we did not
investigate actual information cascades, but we believe that further research is
needed to better elucidate this point.}

We also note how fashion modeling exhibits a strong winner-takes-all component.
In an industry that seems to be governed by such a \emph{survival of the
fittest} mechanism, the difference between performing a show in a premier venue
or not becomes crucial: \rev{while the majority of new faces will not appear in
any prestigious avenue, having even one single runway in one such venue may
decree the success of a new model, bringing her visibility above that of 76\% of
her competitors.\footnote{\rev{Only 24\% of fashion models ran at least one top walk in our dataset.}}} Our analysis aimed at understanding the factors that play a
role in obtaining such popularity, including physical attributes, the reputation
of casting agencies, and the importance of social media presence and reactions.

Regarding our exploratory analysis (\emph{cfr.}~Table~\ref{tab:reg_result}), we
find that thinner and slender individuals are more likely to walk in runways.
Compared to the general population, models are often singled out for their
extremely skinny and tall looks. However, it is interesting that even among
themselves, these preferences --- towards skinny and tall models --- are still
significantly related to the number of runways they can join. \rev{Beauty is
notoriously a hard-to-define quality and, in the case of the fashion industry,
largely a by-product of a collective effort, rather than an inherent
quality~\cite{Mears2011}. While beyond the scope of the present work, an
intriguing question that follows up from it is whether Instagram and other
social media are indeed changing the traditional notions of beauty.}

Research on the fashion industry thus far has been largely qualitative, relying
on methods such as interviews with small number of models and casting
directors~\cite{mears2005not, godart2009cultural}. To the best of our knowledge,
this is first time a large online fashion database has been explored in a
quantitative way, together with data from online social activity. As the impact
of social media --- especially Instagram --- becomes significant in the fashion
industry, predictive methods have the potential to leverage collective attention
and the wisdom of the broader user population, which reflect some of the
popularity of fashion models, to predict their career success. 

\rev{Fashion modeling is one of the best examples of a cultural market, like
music, art, and literature. In all these markets, determining quality of
cultural products is hard because of inherent uncertainty, and thus market
actors must rely on social conventions and buzz as a proxy for success. In
the case of fashion models, here we show that the buzz going on social media
(Instagram in this case) is a reliable predictor of early career success.
Our results are in line with previous work that shows that social signals
have a prominent role in determining success of cultural
contents~\cite{Salganik2006,asur2010predicting}, and so we can expect that
similar approaches to cultural predictions will work in other markets too.
Even scientific production and the stock market are, to some extent, ruled
by prestige and buzz~\cite{Merton1988,Franck1999}. Thus we expect that the
essence of our findings might inform cultural producers and scholars well
beyond the mere fashion industry.}

\rev{In conclusion, computer-mediated collectives are increasingly disrupting
the way culture is consumed and produced. Understanding how use of internet
communication platforms affects cultural production is just an instance of
the study of work in computer-mediated environments and an interesting challenge
for future research.}

\section{Acknowledgements}


The authors would like to thank Bria Carter, Chela Blunt, and Johnny Villamil
for their help with data coding; Maureen Briggs, Alessandro Flammini, and
Filippo Menczer for useful feedback. GLC was supported in part by the Swiss
National Science Foundation (fellowship no. 142353) and the NSF (grant
CCF-1101743). EF acknowledges the support by DARPA grant W911NF-12-1-0034.

\balance \bibliographystyle{SIGCHI-Reference-Format}
\bibliography{jaehyuk_fashion}

\end{document}